\documentstyle [prl,aps,epsfig,multicol]{revtex}
\begin{document}                                                    
\draft                                                      
\title{
Phase transition and phase diagram at a general filling in the
spinless one-dimensional Holstein Model}
\author{Sanjoy Datta  and Sudhakar Yarlagadda}
\address{ Saha Institute of Nuclear Physics, Calcutta, India}
\date{\today}
\maketitle

\begin{abstract}
Among the mechanisms for lattice structural deformation,
the electron-phonon interaction mediated Peierls charge-density-wave (CDW) 
instability in single band low-dimensional systems is perhaps the most
ubiquitous. The standard mean-field picture predicts that the CDW transition
occurs at all fillings and all values of the electron-phonon coupling
$g$ and the adiabaticity parameter $t/\omega_0$.
Here, we correct the mean-field expression for the Peierls instability
condition by showing that the non-interacting static susceptibility, 
 at twice the Fermi momentum, should be replaced by the dynamic one.
We derive the Luttinger liquid (LL) to CDW transition condition,
{\it exact to second order in a novel blocked perturbative approach},
 for the spinless one-dimensional Holstein model in the adiabatic regime.
 The small parameter is the ratio $g \omega_0/t$.
We present the phase diagram at non-half-filling by obtaining
the surprising result that the CDW occurs in a more restrictive region of 
a two parameter ($g^2 \omega_0/t$ and $t/\omega_0$) space than at half-filling.
\end{abstract}

\nopagebreak
\begin{multicols}{2}
\vspace{1cm}

\section{INTRODUCTION}
Over the last few decades, electron-phonon interaction physics has offered
a variety of intriguing and exciting phenomena
such as superconductivity (inorganic and organic), CDW states,
 colossal magnetoresistance, metal-insulator transition,
polaronic ordered phases, etc \cite{alex}. 
Of the electron-phonon models available, the spinless Holstein model
\cite{holstein,fehske0} is a simple and widely used model that mimics   
strongly correlated electron systems with  a strong on-site coulombic
repulsion and short range electron-phonon interactions.
It was conjectured a long while ago by Peierls that 1D electron-phonon
metallic systems, such as the spinless Holstein model, will undergo an
electronic charge-density-wave (CDW) transition with a concomitant
lattice distortion of the same periodicity \cite{peierls}.
Quasi-1D organic charge transfer salts [such as TTF(TCNQ)] and
 conjugated polymers  [such as ${\rm  (CH)_{x}}$] as well as 
 inorganic blue bronzes (e.g., ${\rm K_{0.3}MoO_3}$), 
mixed valence Platinum chain compounds (e.g., Krogmann's salt),
and transition metal chalcogenides (e.g., ${\rm Nb Se_3}$)
\cite{yamaji,tsuda,pierre,gruner} exhibit such 
spontaneous symmetry breaking in the ground state and
 are good candidates for the Holstein model. Furthermore, even the 
strongly correlated  two-band manganite  systems
(e.g., ${\rm La_{1-x}Ca_x MnO_3}$) \cite{cnr}, in the low-doped regime,
can be modeled using the Holstein model \cite{dattays,tvr}. 

Contrary to the mean-field picture, at half-filling, it is now clear that 
the LL to CDW transition occurs only above a critical electron-phonon
coupling strength that depends on the adiabaticity.
Progress has been made over the last few decades, in terms of studying
the Peierls transition at half-filling of the spinless
Holstein model, by using various techniques such as
quantum Monte Carlo simulations \cite{fradkin1,fradkin2,hamer0,capone},
two-cutoff renormalization-group analysis 
\cite{caron}, variational method \cite{zheng1,perroni},
 density-matrix renormalization group (DMRG) method \cite{hamer},
and exact diagonalization \cite{fehske}. However,
a controlled analytic treatment of the quantum phase transition 
has been reported only recently and that too only in the anti-adiabatic
regime \cite{sdadys}. Contrastingly only little effort has been devoted 
to understand the quantum phase transition away from
half-filling \cite{fehske3}.

The present paper is aimed at
providing a well controlled analytic approach to understand the
Peierls quantum phase transition in the adiabatic regime for
the one-dimensional spinless Holstein model at a general filling.
We employ a novel blocking approach that avoids the difficulties
posed by both time-dependent- and degenerate-perturbation theories.
Using the condition that the effective phonon frequency becomes
soft at the symmetry breaking point, we obtain an instability 
criterion that preempts the energy levels crossing condition
for phase transition. In the adiabatic regime and at half-filling,
 we capture the essential features of the LL-CDW transition results
obtained by the ``benchmark" DMRG method in Ref. \cite{hamer}.
At fillings other than half, we show that the LL phase certainly exists
 in the small polaron limit when $g^2\omega_0 /t >> {\rm max}(1,t/\omega_0)$
and in the extreme anti-adiabatic regime. Furthermore, in the adiabatic
regime, we also demonstrate that CDW phase does exist at
intermediate values of the electron-phonon couplings $g <t/\omega_0$.
 We propose  a qualitative phase diagram as a guide for future work.

\section{PHONON SOFTENING IN THE LL PHASE}
 We begin by considering the phonon-softening in the LL phase
as the signal for lattice deformation in the 1D Holstein model. 
The non-interacting Hamiltonian 
\begin{eqnarray}
H_0 = 
 \sum _{{\vec{k}}^{\prime}} \epsilon_{{\vec{k}}^{\prime}}
  c _{{\vec{k}}^{\prime}}^{\dagger}  c _{{\vec{k}}^{\prime} } 
+ \omega_0 \sum_{\vec{q}} 
a_{\vec{q}}^{\dagger} a_{\vec{q}} ,
\label{H0}
\end{eqnarray}
and the perturbation 
\begin{eqnarray}
H_{1} = 
\frac{g \omega_0}{\sqrt{N}} 
 \sum_{ \vec{q} }
\rho_{\vec{q}}
( a_{\vec{q}} +
a_{-\vec{q}}^{\dagger}) ,
\label{H1}
\end{eqnarray}
together make up the Holstein Hamiltonian. In the above equations,
$ \rho_{\vec{q}} = 
 \sum _{{\vec{k}}^{\prime}}
c^{\dagger}_{{\vec{k}}^{\prime} +\vec{q}}
  c _{{\vec{k}}^{\prime} }$ 
 is the density operator,
  $c _{{\vec{k}}^{\prime} }$ 
is  the
 electron  destruction operator with
  ${\vec{k}}^{\prime} $ 
 limited 
to the first Brillouin zone,
 $ \epsilon_{{\vec{k}}^{\prime}} =-2t \cos(k^{\prime})$
 with lattice constant
being taken to be unity, $t$ is the hopping integral,
$ a_{\vec{q}}$ is the phonon destruction operator,
$\omega_{0}$ is the optical phonon frequency,
 and $N$ is the number of sites.
The eigen states and eigen energies of $H_0$ 
are given by $|\phi_l \rangle = |n;m \rangle \equiv 
|n \rangle_{el} |m\rangle_{ph}$
(with $|\phi_0 \rangle = |0;0\rangle$ being the ground state with 
zero phonons) and $E^0_{\phi_l}$ respectively. Whereas, for the
interacting Hamiltonian $H=H_0 + H_1$, the corresponding 
eigen states are $|\Phi_l \rangle$ and the eigen energies are $E_{\Phi_l}$.

Now the double time derivative of an operator $A$ is given by
 \begin{equation} 
\ddot{A} = - [[A, H], H ] .
\label{Q_tt}
 \end{equation} 
From the above equation, when $A$ is taken to be the 
ionic position coordinate $Q_{\vec{p}}
=\sqrt{1/(2M\omega_0)} (a_{\vec{p}} + a_{-\vec{p}}^{\dagger})$ 
and upon making the static mean-field approximation 
$ \rho_ {-\vec{p}} \propto \chi_0 (\vec{p},0) Q_{\vec{p}}$
\cite{gruner}, we obtain the following expression:
 \begin{equation} 
\ddot{Q}_{\vec{p}} =
- \omega _0^2 
 [1+2 g^2 \omega_0  \chi_0 (\vec{p}, 0) ] 
Q_{\vec{p}} 
=- \omega_R^2
Q_{\vec{p}} .
\label{stdinst}
 \end{equation} 
We know that in 1D the non-interacting polarizability
 $\chi_0 (\vec{p},0)$ has a negative divergence at wavevector $p=2k_F$.
Thus it appears that the renormalized phonon frequency $\omega_R$ becomes
soft even for vanishingly small electron-phonon interaction leading to
lattice distortion. We will now proceed to derive the true phonon 
softening condition rigorously. To this end we calculate the matrix elements 
of Eq. (\ref{Q_tt}) and obtain  
 \begin{equation} 
\langle \Phi_m | \ddot{A} |\Phi_n \rangle  = -
(E_{\Phi_m} - E_{\Phi_n})^2 \langle \Phi_m | A |\Phi_n \rangle . 
\label{A_matrix}
 \end{equation} 
 When $\omega_{e}^2 =(E_{\Phi_m} - E_{\Phi_n})^2 \le 0$,
instability occurs for transition from $|\Phi_n \rangle $ to 
$|\Phi_m \rangle$   provided that 
$\langle \Phi_m  | A |\Phi_n \rangle \neq 0  $. For the total Hamiltonian
 $H_0 + \lambda H_1$, we obtain $E_{\Phi_n}$ perturbatively to be 
$E_{\Phi_n}= E_{\phi_n}^0 + \lambda^2 E_{\phi_n}^{(2)} + ...$
where $ E_{\phi_n}^{(2)}$ is the second order correction to the energy.
Then, to second order in $\lambda$, we obtain
 \begin{eqnarray} 
\!\!\! \omega_{e}^2 = 
(E_{\phi_m}^0 - E_{\phi_n}^0)^2 
+ 2 \lambda ^2 
(E_{\phi_m}^0 - E_{\phi_n}^0) 
(E_{\phi_m}^{(2)} - E_{\phi_n}^{(2)}) . 
\label{Enm2}
 \end{eqnarray} 
Thus to order $\lambda ^2$ in perturbation, 
as the strength of the interaction is increased,
$\omega_{e}^2 = 0$ before $E_{\Phi_m} = E_{\Phi_n}$. 

We will now focus on the eigen states $\Phi_l$ and the eigen energies
$E_{\Phi_l}$ of $H_0 + H_1$ to obtain the phonon softening condition.
The energy of the (expected) ground state, upto second order in perturbation,
 is given by:
 \begin{eqnarray} 
E_{\Phi_0} = T_0 
-\frac{g^2 \omega_0^2}{N} 
\sum_{ \vec{q}, m \neq 0 } 
\frac{|\langle m|_{el} \rho_{\vec{q}} |0\rangle _{el}|^2}
{ \xi_{m0}+ \omega_0} ,
\label{Ephi0}
 \end{eqnarray} 
where $T_0$ ( = $E^0_{\phi_0}$) is
 the non-interacting kinetic energy of the $|\phi_0 \rangle$ state
 and $\xi_{m0} \equiv \xi_m - \xi_0$
with $\xi_m$ being the energy of $|m \rangle _{el}$. 
Let $|\psi_n^0 \rangle \equiv |0 ; n_{-\vec{p}} \rangle $ with
$|n_{-\vec{p}} \rangle_{ph} $ corresponding to a state with $n$ phonons all
of which being in the $-\vec{p}$ state. Then, the corresponding 
interacting state $|\Psi_n^0 \rangle $ yields the energy difference 
$E_{\Psi^0_{n+1}} -E_{\Psi^0_{n}} = \omega_0 + 
\Sigma (-\vec{p} , \omega_0 )$ where
the self-energy, which is complex in general, is given by 
$\Sigma (\vec{p} , \omega_0 ) = g^2 \omega_0^2 \chi_0 (\vec{p} , \omega_0 )$
[see Appendix A for details]
with the non-interacting (Lindhard) polarizability
$\chi_0$ being defined as \cite{lindhard} 
 \begin{eqnarray*} 
\chi_0 (\vec{p}, \omega_0) \equiv 
 \frac{1}{N} 
 \sum_{m \neq 0} 
 \left [
\frac{|\langle m|_{el} \rho_{\vec{p}} |0\rangle _{el}|^2}
{\omega_0-\xi_{m0}+i\eta}
 -
\frac{|\langle m|_{el} \rho_{-\vec{p}} |0\rangle _{el}|^2}
{\omega_0+\xi_{m0}+i\eta} \right ] .
 \end{eqnarray*} 
Hence we see that, although the above energy difference 
yields the expression
$\omega_{e}^2 = \omega_0^2 [1+ 2 g^2 \omega_0 \chi_0 (\vec{p} , \omega_0 )]$
[based on  Eqs. (\ref{A_matrix}) and (\ref{Enm2})]
 with the form being similar to that in Eq. (\ref{stdinst}), 
the complex nature of the self-energy complicates 
identifying the phonon softening condition for lattice instability
\cite{chi0}. We will adopt an alternate perturbative procedure to 
obtain the lattice instability criterion.

\section{HALF-FILLED CASE}
We begin by observing that the expression
 \begin{eqnarray} 
\omega_0 = \epsilon_{\vec{k}-\vec{p}} -\epsilon_{\vec{k}} = 
4t \sin(k_F - k) \sin(k_F) ,
\label{degcond}
\end{eqnarray} 
with  $p = 2k_F$ has two solutions for $k$.
Then at half-filling ($k_F = \pi/2$) and $k < k_F$,
the two solutions to Eq. (\ref{degcond})
are $\pm k = \pi/2 - \arcsin [\omega_0 /4 t ]$. Thus,  the states
 $|\psi_1^1 \rangle \equiv |\vec{k} \rightarrow \vec{k}-\vec{p} ; 0 \rangle $
(corresponding to exciting, from the ground state,
the electron at $\vec{k}$
to $\vec{k}-\vec{p}$) and $|\psi_1^2 \rangle \equiv 
|-\vec{k} \rightarrow -\vec{k}-\vec{p} ; 0 \rangle $ 
are degenerate with
$|\psi_1^0 \rangle \equiv |0 , 1_{-\vec{p}} \rangle $ and are connected
to $|\psi_1^0 \rangle $ through $H_1$ as
$\langle \psi_1^{1,2} |H_1 |\psi_1^0 \rangle \neq 0$.
Hence,  employing degenerate perturbation theory seems to be a natural
choice to study lattice period doubling.
 However, the number of degenerate states that
need to be considered increases linearly
with number of phonons in the state
$|\psi_n^0 \rangle \equiv |0 ; n_{-\vec{p}} \rangle $
[see Appendix B.1].
Then, to calculate $\omega_e^2$ for large $n$ becomes difficult!
\begin{figure}[b]
\includegraphics[width=3.0in]{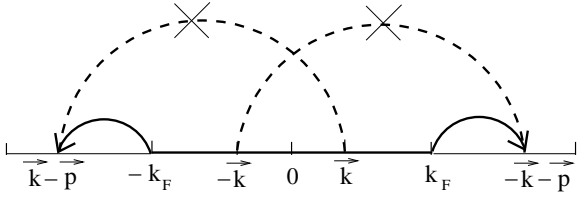}
\vspace*{0.5cm}
\noindent \caption[]{{\bf Blocking procedure at half-filling.} 
Electrons at $\mp k_F=\mp \pi/2 $ are excited from the ground state
(dark line) to $\pm \vec{k} - \vec{p}$ to block the excitations
 $\pm \vec{k} \rightarrow \pm \vec{k} - \vec{p}$ respectively.}
\label{block}
\end{figure}
To circumvent the above problem, we adopt the following approach.
We choose our starting state to be
$|\phi_n \rangle \equiv |-\vec{k_F} \rightarrow \vec{k}-\vec{p} ,
\vec{k_F} \rightarrow -\vec{k}-\vec{p} ; n_{-\vec{p}}  \rangle $ ($n \ge 1$)
such that, by exciting the electrons at $\mp \vec{k}_F$
from the ground state 
 to $\pm \vec{k}-\vec{p}$ outside the Fermi sea (FS), the  
excitations $\vec{k} \rightarrow \vec{k}-\vec{p} $
and $-\vec{k} \rightarrow -\vec{k}-\vec{p}$ are Pauli blocked [as shown in 
Fig. \ref{block}]. Now, $\pm \vec{k_F}$ have been excited for
ease of mathematical manipulation.
The state $\phi_n$, upon turning on interactions, yields the energy
[see Appendix B.2 for details]:
 \begin{eqnarray} 
E_{\Phi_n}=
&&
E_{\Phi_0}+ 
 (n+1)\omega_0 [1+ g^2 \omega_0 {\rm Re}
 \chi_0 (\vec{p}, \omega_0)]
 \nonumber \\
 &&
 + g^2 \omega_0^2
 {\rm Re} [2 \Pi_0 (\vec{k}-\vec{p}, \vec{k}_F, \omega_0)
- \chi_0 (\vec{p}, \omega_0)] ,
\label{Ephin2}
\end{eqnarray}
where
 \begin{eqnarray*} 
{\rm Re} \Pi_0 (\vec{s}, \vec{q}, \omega_0) 
\equiv 
- \frac{1}{N}
 && \sum_{\vec{r}} 
\left [ 
\frac{1-n_{\vec{r}}}
{
\epsilon_{\vec{r}} + \omega_0 -
\epsilon_{\vec{s}}}
+
\frac{n_{\vec{r}}}
{
\epsilon_{\vec{q}} + \omega_0 -
\epsilon_{\vec{r}}
}
\right .
\nonumber \\
&& ~
\left .
 -
\frac{n_{\vec{r}}}
{
\epsilon_{\vec{s}} + \omega_0 -
\epsilon_{\vec{r}}
}
-
\frac{1-n_{\vec{r}}}
{
\epsilon_{\vec{r}} + \omega_0 -
\epsilon_{\vec{q}}
}
 \right ] .
 \end{eqnarray*} 
For $\omega_0/(4t) << 1$,
 \begin{eqnarray} 
 {\rm Re} [2 \Pi_0 (\vec{k}-\vec{p}, \vec{k}_F, \omega_0)
- \chi_0 (\vec{p}, \omega_0)] 
\approx
  \frac{1}{2 \pi t} \ln \left [ \frac{8 t}{9 \omega_0 } \right ] .
\end{eqnarray}
The above approximation underestimates the actual value of $2 \Pi_0 -\chi_0$
by less than 5\% for $t/\omega_0 > 4$.
In Appendix C,  Fig. 6(c) shows that $2\Pi_0 - \chi_0 > 0$
 for $t/\omega_0 > 1$.
In the above Eq. (\ref{Ephin2}), for $n \rightarrow \infty$, energy instability
($ E_{\Phi_{n}} - E_{\Phi_0} < 0$)  
occurs for values of $g$ larger than $g_E$ given by 
 $1+ g^2_E \omega_0 {\rm Re} \chi_0 (\vec{p}, \omega_0)=0$.
For $g > g_E$,  $E_n$ {\em has no lower bound} which is an unphysical
situation.  Also when $g> g_E$,
$ E_{\Phi_{n+1}} - E_{\Phi_{n}} =  
 1+ g^2 \omega_0 {\rm Re} \chi_0 (\vec{p}, \omega_0)< 0 $ for all $n\ge1$
which leads to the remarkable situation that all $E_{\Phi_n}$ cross
at the same $g=g_E$.
To second order in the small parameter of perturbation, 
 similar to Eq.(\ref{Enm2}), one obtains
\begin{eqnarray}
 \omega_{e}^2
= 
 (E_{\Phi_n} - E_{\Phi_0} )^2
 \approx
 &&
 (n+1)^2 \omega_0^2 [1+2 g^2 \omega_0 {\rm Re}
 \chi_0 (\vec{p}, \omega_0) ]
 \nonumber \\
 &&
 +2(n+1) \frac{g^2 \omega_0^3}{2 \pi t} \ln \left [ \frac{8 t}{9 \omega_0 }
\right ] .
\label{weff}
\end{eqnarray}
Thus we see from Eq. (\ref{weff}) that, in the adiabatic regime
and for large $n$, the above mentioned
 energy instability occurring at $g > g_E$ is pre-empted by the
phonon softening occurring at $g > g_c = g_E/\sqrt{2}$  
 with $g_c$ defined by the following expression:
\begin{eqnarray}
1+ 2 g^2_c \omega_0 {\rm Re}
 \chi_0 (\vec{p}, \omega_0)=0 .
\label{gcchi0}
\end{eqnarray}
The above equation is one of our main results and is the correction
to the mean-field instability condition 
$1+ 2 g^2 \omega_0 \chi_0 (\vec{p}, 0)=0 $ obtained from
 Eq. (\ref{stdinst}). The operator $A$, which
produces non-vanishing matrix elements in Eq. (\ref{A_matrix}), is given by
 $A = c^{\dagger}_{\vec{k} - \vec{p}} c_{- \pi/2}
c^{\dagger}_{-\vec{k} - \vec{p}} c_{ \pi/2} (a^{\dagger}_{-\vec{p}})^n$.
Thus, the system becomes
unstable towards absorbing a large number ($n$) of phonons leading to a
macroscopic deformation as explained below.
The displacement-displacement correlation function is given by 
\begin{eqnarray}
\langle \Phi_n | Q_l Q_j | \Phi_n \rangle =
\langle \phi_n | Q_l Q_j | \phi_n \rangle 
= 
\frac{n \cos [(j-l)p]}{N M \omega_0} .
\label{modul}
\end{eqnarray}
Thus we see that, for non-vanishing values
of $n/N$, one obtains an observable ionic-position modulation when 
$|\Phi_n \rangle$ is the interacting ground state.
The above Eq. (\ref{modul}) is true for all fillings 
and for any eigen state with $n_{-\vec{p}}$ phonons.

\begin{figure}[t]
\includegraphics[width=3.0in]{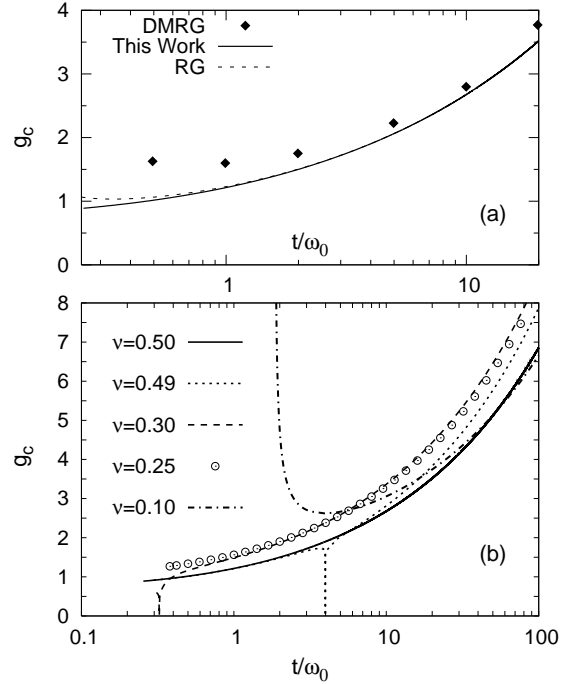}
\vspace*{0.5cm}
\noindent\caption[]{{\bf Critical coupling $g_c$ versus adiabaticity parameter
$t/\omega_0$.} Comparison of the $g_c$ values obtained at
(a) half-filling in this work, 
in Ref. \cite{hamer} using DMRG, and in Ref. \cite{caron} using two-cutoff RG;
and (b) various fillings ($\nu$) in this work.}
\label{mlptgc}
\end{figure}

The critical coupling $g_c$, given by Eq. (\ref{gcchi0}),
can be expressed analytically as follows:
 \begin{eqnarray} 
\frac{\pi}{g_c^2} = \frac{\gamma}{\sqrt{1 - \gamma ^2 }}
 && \left \{ 
\ln \left [ \frac
{( 1 - \sqrt{1 - \gamma^2})^2 - (\gamma \tan k_F )^2}
{( 1 + \sqrt{1 - \gamma^2})^2 - (\gamma \tan k_F )^2} \right ] \right .
\nonumber \\
&&
~~~
\left .
- 2 \ln \left [ \frac
{ 1 - \sqrt{1 - \gamma^2} }
{ 1 + \sqrt{1 - \gamma^2} } \right ] \right \} ,
\label{instabgc}
 \end{eqnarray} 
where $\gamma \equiv \omega_0/(4t \sin(k_F)) < 1$.
Fig. \ref{mlptgc} depicts, for various
filling factors $\nu$, the variation of the critical coupling
 $g_c$ with the adiabaticity parameter $t/\omega_0$.
At half-filling, for values of $t/\omega_0 > 5$,
 our theoretical curve is quite close to the numerically
determined values of $g_c$ as reported in Ref. \cite{hamer}.
Furthermore, at half-filling and for $\gamma^2 << 1$,
our expression for $g_c$ [given by Eq. (\ref{instabgc})]
reduces to the two-cutoff 
renormalization result of Caron and Bourbonnais \cite{caron},
i.e., $\omega_0 = 2 c t \exp(-\pi t /g^2 \omega_0)$ (with $c \sim 1$), 
when we take $c=4$. The numerical agreement between the two expressions
is depicted in Fig. \ref{mlptgc}(a).

\section{LESS THAN HALF-FILLING}
\begin{figure}[b]
\includegraphics[width=3.0in]{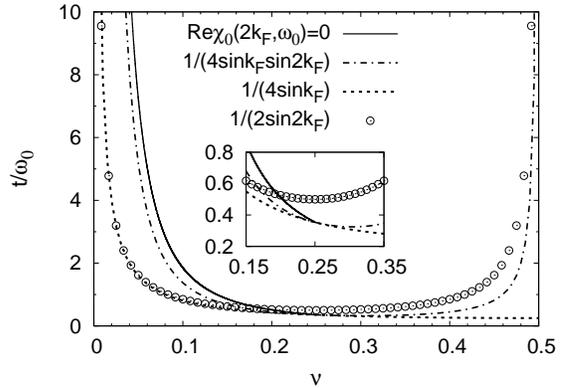}
\vspace*{0.5cm}
\noindent \caption[]{{\bf Curves relevant for identifying
different perturbative
regimes.} Plot of $t/\omega_0$ and filling factor $\nu =k_F/\pi$ values
 satisfying ${\rm Re}\chi_0 (2 k_F , \omega_0 ) =0$ and curves for
the functions $t/\omega_0 = 1/[4 \sin (k_F) \sin (2 k_F)]$, 
$= 1/[4 \sin (k_F)]$, and $= 1/[2 \sin (2 k_F)]$. 
}
\label{chiff}
\end{figure}

We will now consider fillings that are less than half-filling.
The line depicted by $t/\omega_0  = 1/[4 \sin(2k_F) \sin(k_F)]$ 
[obtained by setting $k=-k_F$ in Eq. (\ref{degcond})], 
corresponds to the divergence of $ \chi_0 (\vec{p}, \omega_0)$.
 In the region above $t/\omega_0  = 1/[4 \sin(2k_F) \sin(k_F)]$
(see Fig. \ref{chiff}), the excitation energy expression
$ \epsilon_{\vec{k}-\vec{p}} - \epsilon_{\vec{k}} = \omega_0$ 
with $p = 2 k_F$ is satisfied by one  wavevector for $|k| < k_F$
 [as seen from Eq. (\ref{degcond})] (see Appendix C
 for a complete analysis). We will now consider
 the region $t/\omega_0  >  1/[4 \sin(2k_F) \sin(k_F)]$.
In this region, the degenerate states are
$|\psi_n^0 \rangle \equiv |0 ; n_{-\vec{p}} \rangle $ and $|\psi_n^1 \rangle
 \equiv |\vec{k} \rightarrow \vec{k}-\vec{p} ; (n-1)_{-\vec{p}} \rangle $.
We obtain the lower eigen energy, which corresponds to the state
$|\psi_n^- \rangle \equiv [|\psi_n^0 \rangle - |\psi_n^1 \rangle]/\sqrt{2}$, 
 to be (see Appendix C)
 \begin{eqnarray} 
E_{\Psi_n^-}=
&&
E_{\Phi_0}+
E^1
 +n\omega_0 [1+ g^2 \omega_0 {\rm Re}
 \chi_0 (\vec{p}, \omega_0)]
\nonumber
\\
&&
 - 0.5 g^2 \omega_0^2 {\rm Re}
 [\chi_0 (\vec{p}, \omega_0) - \Pi_0 (\vec{k}-\vec{p}, \vec{k}, \omega_0)] ,
\label{En1sol}
\end{eqnarray}
 where $E^1 = -\frac{\sqrt{n}}{\sqrt{N}}g\omega_0$ is the first order
energy correction. In arriving at the above energy, we have
ignored the contribution
$n/[N(\epsilon_{\vec{k}-\vec{2p}} -\epsilon_{\vec{k} - \vec{p}} - \omega_0)]$
corresponding to exciting the electron at  
$\vec{k}-\vec{p}$ to the state $ \vec{k} - \vec{2p}$ by
destroying a phonon of momentum $-\vec{p}$. This is valid provided
$\epsilon_{\vec{k}-\vec{2p}} -\epsilon_{\vec{k} - \vec{p}} - \omega_0 \neq 0$.
The case when
$\epsilon_{\vec{k}-\vec{2p}} -\epsilon_{\vec{k} - \vec{p}} - \omega_0 = 0$
(i.e., $t/\omega_0 = 1/[2 \sin (2 k_F)]$),  
will be discussed in the Appendix C.1.
Obviously, the macroscopic deformation instability condition
 is still $g>g_c$ with $g_c$ given by  Eq. (\ref{gcchi0}).

It should be pointed out that only the region
$ {\rm Re} \chi_0(\vec{p},\omega_0) < 0$ is relevant in obtaining $g_c$.
The curve $ {\rm Re} \chi_0(\vec{p},\omega_0) = 0$ is depicted
in Fig. \ref{chiff}
 and exists only for $k_F < \pi/4$. Above (below) this curve,
$ {\rm Re} \chi_0(\vec{p},\omega_0) $ is always negative (positive).
Furthermore, for $k_F > \pi/4$ and all values of the adiabaticity
parameter $t/\omega_0$ above (below) the line
 $ t/\omega_0 = 1/[4 \sin(k_F)] $, one can show analytically
 that $ {\rm Re} \chi_0(\vec{p},\omega_0) $ is always negative (positive).

It can be shown that, for a given filling $\nu$ and any value of $t/\omega_0$
where $ {\rm Re} \chi_0(\vec{p},\omega_0) < 0$, the macroscopic
instability condition is always given by Eq. (\ref{gcchi0}) 
(see Appendix C). 
For filling factors above $0.25$, as shown in Fig. \ref{mlptgc},
 the $g_c$ decreases with decreasing
$t/\omega_0$ with a downward kink appearing at a certain value of $t/\omega_0$
corresponding to the negative divergence of $ \chi_0(\vec{p},\omega_0)$.
At the point of divergence of $ \chi_0(\vec{p},\omega_0)$, perturbation theory
is no longer valid.
For $k_F < \pi/4$, the $g_c$ initially decreases with decreasing $t/\omega_0$
until a certain value of $t/\omega_0$; while below this value
$t/\omega_0$, the value of $g_c$ again increases due to the fact that 
$ {\rm Re} \chi_0(\vec{p},\omega_0) $ value approaches zero value 
(see Fig. \ref{chiff}).
Lastly, we would like to point out that the $g_c$ values are not reliable
when $g_c \omega_0/t > 1$ and hence 
in the entire anti-adiabatic regime ($t/\omega_0 < 1$) the $g_c$ values are
suspect [see Fig. \ref{mlptgc}(b)].

\section{T=0 PHASE DIAGRAM AT NON-HALF-FILLING}
At non-half filling, we will now discuss the quantum phase transition 
based on the perturbation
theory results derived above and the work reported in Ref. \cite{sdadys}.
In the extreme small polaron regime, for $g > 1$, it was shown earlier
that  the effective Holstein Hamiltonian
 can be recast as an effective spin Hamiltonian (using Wigner-Jordan 
transformation)  as follows (see \cite{sdadys} for details):
\begin{eqnarray}
H^{spin}_{e} \sim   -g^2 \omega_0 && \left [
  \sum_j \sigma^{z}_j
 + \zeta e^{-g^2} \sum_j (
\sigma^{+}_{j}\sigma^{-}_{j+1} 
 + {\rm H.c.}) \right . 
\nonumber \\
&&
 - \zeta ^2 \sum_j 
\sigma^{z}_j \sigma^{z}_{j+1} 
\nonumber \\
&&
\left .
+ \zeta ^2 e^{-g^2} \sum_j  \{ 
\sigma^{+}_{j-1} \sigma^{-}_{j+1}
 + {\rm H.c.} \}
 \right ] ,
\label{Heff_sp}
 \end{eqnarray}                               
where $\zeta \equiv t/g^2 \omega_0$ is the polaron size parameter.
The above equation was obtained by
assuming that the phonons are frozen in the
Lang-Firsov transformed (LFT)
phononic ground state $\exp [g \Sigma_j c^{\dagger}_j c_j (a_j - a_{j+1})]
|0 \rangle _{ph}$ \cite{lang}.
Now, for the above Eq. (\ref{Heff_sp}) to be the basis for
studying phase transition, each of the coefficients of 
the second, third, and fourth terms on the rhs should be significantly smaller
than $\omega_0$ (so that the LFT phononic ground state remains unaffected).
In Eq. (\ref{Heff_sp}), for $\zeta << 1$, the coefficients of
the nearest neighbor and the next to nearest neighbor interactions
in the transverse direction are much smaller than $\omega_0$. 
Contrastingly, the coefficient of the nearest neighbor interaction
in the longitudinal direction is much smaller
than $\omega_0$ always when $t/\omega_0 < 1$;  
while for $t/ \omega_0 > 1$, it is much smaller
only when $[t/(g \omega_0)]^2 << 1$ (i.e., for large values of $g$).
Note that, when $\zeta << 1$, the last term is negligible.
Using Bethe ansatz, we know that anisotropic Heisenberg model always yields
a Luttinger liquid away from half-filling \cite{ibose,giamarchi,haldane}!
From the above analysis, it follows that a LL results 
for all values of $1/\zeta >> {\rm max} (1, t/\omega_0 )$.
Furthermore, for $t/\omega_0 << 1$ and $g > 1$, the above Eq. (\ref{Heff_sp})
is valid with the last term on the rhs being negligible and 
consequently LL results away from half-filling.

Next, when $g < 1$ and $t/\omega_0 << 1$, we get the corresponding
effective spin Hamiltonian from the effective Holstein Hamiltonian
to be \cite{sdadys}  
\begin{eqnarray}
H^{spin}_{e} \sim  -g^2 \omega_0  && \left [  
  \sum_j \sigma^{z}_j
 + \zeta e^{-g^2} \sum_j (
\sigma^{+}_{j}\sigma^{-}_{j+1} 
 + {\rm H.c.})  \right .
\nonumber \\
&&
+ 
 \left ( {\frac{t}{\omega_0}} \right ) ^2 e^{-2 g^2} 
 \sum_j  \{ 
\sigma^{+}_{j-1} \sigma^{-}_{j+1}
+ {\rm H.c.} \} 
\nonumber \\
&&
\left .
  -
4 \left ( {\frac{t}{\omega_0}} \right ) ^2 e^{-2 g^2} 
 \sum_j 
\sigma^{z}_j \sigma^{z}_{j+1} 
\right ]
 .
\label{Heff2}
 \end{eqnarray}                              
In arriving at the above equation too, it was assumed that the phonons
are in the LFT phononic ground state. Such an assumption is justified because
the coefficients of the second, third, and fourth terms on the rhs
of the above equation, are much smaller than $\omega_0$.
In Eq. (\ref{Heff2}), the small parameter is $t/\omega_0$ and the last 
two terms are negligible compared to the second term
when the adiabaticity parameter $t/\omega_0 << 1$. Then, 
this implies a LL state for all fillings. For $g \rightarrow 0$ and 
any value of $t/\omega_0 $, we do not expect a CDW state.

\begin{figure}[t]
\includegraphics[width=3.0in]{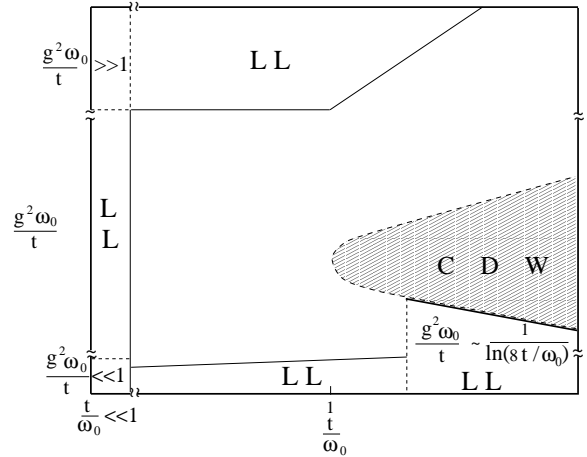}
\vspace*{0.5cm}
\noindent \caption[]{
{\bf Zero temperature phase diagram at non-half filling.}
Regions where CDW and LL phases certainly exist are depicted.
 The calculated transition from LL to CDW is indicated by a dark line
along with its expression.}
\label{phasediag}
\end{figure}

The phase diagram (see Fig. \ref{phasediag}) is drawn qualitatively
for a general filling away from half-filling.
For $\nu < 0.25$, the CDW region shifts to the right with decreasing $\nu$
as can be surmised from the region of validity $g_c \omega_0 /t < 1$
in Fig. \ref{mlptgc}(b).  The regions where LL is certain
is  indicated. For $t/\omega_0 > 1$, since we need  
 $[t/(g \omega_0)]^2 << 1$ for the validity of Eq. (\ref{Heff_sp}),
 the boundary of the LL-certain-region is linear and of the form
 $g^2 \omega_0/t = D t/\omega_0$ where the slope $D >> 1$.
 Furthermore, for $g \omega_0/t << 1$,
one expects a LL phase and hence we get a linear boundary
(of the form $g^2 \omega_0/t = d t/\omega_0$ with $d << 1$)
 for the LL phase in the lower left part of the diagram.

The thatched portion corresponds to a region where a CDW state is certain. 
The upper boundary of the thatched region,
which is obtained by the condition $g \omega_0 /t \sim 1$
and is therefore linear with slope of order unity,
corresponds to the breakdown of the perturbation theory used in identifying
the CDW transition. The only certain location of the transition
from LL to CDW is indicated by a solid dark line and is
approximately given by
$g^2\omega_0/t \sim 1/(\ln(8t/\omega_0)$ [while the exact relation is
expressed in Eq. (\ref{instabgc})].
However, it is unclear in the rest of the figure where exactly the transition
from a CDW  state to a LL state takes place.

Finally,  it should be emphasized that there are two different types of
phase transitions. The nature of the phase transition
on the adiabatic side, at intermediate values of $g < t/\omega_0$,
 is driven by a macroscopic ionic lattice distortion. Quite differently, 
in the restricted region of the small polaron limit
where $1/\zeta >> {\rm max} (1, t/\omega_0 )$ and in the extreme anti-adiabatic
regime ($t/\omega_0 << 1$), the CDW is driven by
  a small-polaron-interaction based mechanism.
In the latter case, the coordinate Bethe ansatz argument 
precludes the possibility of a small polaronic CDW
away from half-filling in sharp contrast to the half-filled case 
\cite{hamer,sdadys}.

\section{ACKNOWLEDGEMENTS}
One of the authors (S.Y.) would like to thank G. Baskaran and 
S. M. Bhattacharjee for useful discussions. This work was partially funded
by CAMACS of SINP.

\end{multicols}

\pagebreak

\appendix
\section{}
\begin{figure}
\begin{center}
\includegraphics[width=3.0in]{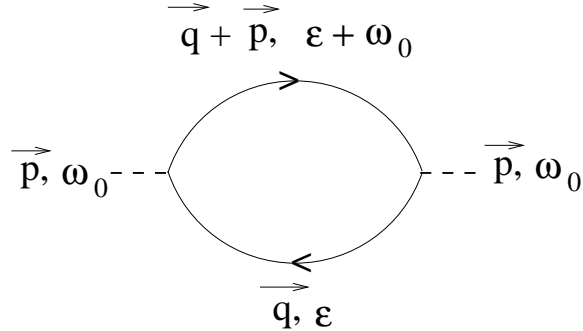}
\end{center}
\caption[]{ Self-energy of a phonon. The solid and the dashed
lines depict the electron and the phonon propagators respectively.}
\label{selfeng}
\end{figure}

The interacting state $|\Psi_n^0 \rangle$, corresponding to the
non-interacting state 
$|\psi_n ^0 \rangle= |0; n _{-\vec{p}} \rangle$, yields the following energy
expression: 
 \begin{eqnarray}
E_{\Psi_n^0} = T_0+ n \omega_0
-\sum_{\phi_l\neq \psi_n^0}
 \left [ 
\frac{|\langle \phi_l| H_1 |\psi_n^0\rangle |^2}
{E_{\phi_l}^0 - E_{\psi_n^0}^0}
 \right ] ,
\label{Epsin0}
\end{eqnarray}
where $T_0 = E^0_{\phi_0} = -2 N t \sin (k_F) /\pi$ is the non-interacting 
kinetic energy of the ground state $|\phi_0 \rangle$ and
 \begin{eqnarray}
\sum_{\phi_l \neq \psi_n^0}
\frac{|\langle \phi_l| H_1 |\psi_n^0\rangle |^2}
{E_{\phi_l}^0 - E_{\psi_n^0}^0}
&&
 =g^2 \omega_0^2
 \left [
\frac{1}{N} 
\sum_{m\neq 0}
\sum_{\vec{q} \neq \vec{p}}
\frac{|\langle m|_{el} \rho_{\vec{q}} |0\rangle _{el}|^2}
{ \xi_{m0}+\omega_0}
+
\frac{n+1}{N} 
\sum_{m\neq 0}
\frac{|\langle m|_{el} \rho_{\vec{p}} |0\rangle _{el}|^2}
{ \xi_{m0}+\omega_0}
\right .
\nonumber \\
&&
~~~~~~~
\left .
+\frac{n}{N} 
\sum_{m\neq 0}
\frac{|\langle m|_{el} \rho_{-\vec{p}} |0\rangle _{el}|^2}
{ \xi_{m0}-\omega_0 - i\eta}
 \right ] 
\nonumber \\
&&
=
 g^2 \omega_0^2
 \left [\frac{1}{N} 
\sum_{ \vec{q}, m\neq 0 }
\frac{|\langle m|_{el} \rho_{\vec{q}} |0\rangle _{el}|^2}
{ \xi_{m0}+\omega_0}
- n \chi_0 (\vec{p}, \omega_0)
 \right ] ,
\label{sec-ord}
\end{eqnarray}
with $\eta \rightarrow 0^+$.
Then the above Eq. (\ref{sec-ord}) yields the energy difference 
$E_{\Psi^0_{n+1}} -E_{\Psi^0_{n}} = \omega_0 + 
\Sigma (-\vec{p} , \omega_0 )$ where
 the self-energy of
 a phonon 
$\Sigma (\vec{p} , \omega_0 ) = g^2 \omega_0^2 \chi_0 (\vec{p} , \omega_0 )$.
The self-energy is displayed in Fig. \ref{selfeng} with the bubble
representing the polarizability 
$ \chi_0 (\vec{p} , \omega_0 )$ and each of the electron-phonon interaction
vertices corresponding to the factor $g \omega_0$. 
\section{}
\subsection{\underline{\textbf{\normalsize{The set of
 degenerate states for half-filled case}}}}
The solutions of the expression
 \begin{eqnarray}
\omega_0 = \epsilon_{\vec{k}-\vec{p}} -\epsilon_{\vec{k}} =
4t \sin(k_F - k) \sin(k_F) ,
\label{degcond2}
\end{eqnarray}
for $p = 2k_F$ are given by
 \begin{eqnarray}
k=k_F-\arcsin \left ( \frac{\omega_0}{4 t \sin (k_F)} \right ) ,
\label{sol1}
\end{eqnarray}
and
 \begin{eqnarray}
k=-\pi + k_F+\arcsin \left ( \frac{\omega_0}{4 t \sin (k_F)} \right ) .
\label{sol2}
\end{eqnarray}
Then, the set of states that are degenerate with $ |0;n_{-\vec{p}} \rangle$ 
and that should be considered in degenerate perturbation theory
consists of the following states  
$|0 ; (n-2m)_{-\vec{p}}
m_{2 \vec{k} - \vec{p}}
m_{-2 \vec{k} - \vec{p}}  \rangle $,
$ |\vec{k} \rightarrow \vec{k}-\vec{p} ; (n-2m-1)_{-\vec{p}}
m_{2 \vec{k} - \vec{p}}
m_{-2 \vec{k} - \vec{p}}  \rangle $,
$ |-\vec{k} \rightarrow -\vec{k}-\vec{p} ; (n-2m-1)_{-\vec{p}}
m_{2 \vec{k} - \vec{p}}  m_{-2 \vec{k} - \vec{p}}  \rangle $,
$ |\vec{k} \rightarrow \vec{k}-\vec{p} ,
-\vec{k} \rightarrow -\vec{k}-\vec{p} ; (n-2m-2)_{-\vec{p}}
m_{2 \vec{k} - \vec{p}}  m_{-2 \vec{k} - \vec{p}}  \rangle $,
$ |\vec{k} \rightarrow -\vec{k}-\vec{p} ; (n-2m-2)_{-\vec{p}}
(m+1)_{2 \vec{k} - \vec{p}}  m_{-2 \vec{k} - \vec{p}}  \rangle $,
$ |-\vec{k} \rightarrow \vec{k}-\vec{p} ; (n-2m-2)_{-\vec{p}}
m_{2 \vec{k} - \vec{p}}
(m+1)_{-2 \vec{k} - \vec{p}}  \rangle $,
where m=0,1,2,3,... with the constraint that the number of phonons is
non-negative. Thus we see that the number of degenerate states 
increases linearly with $n$ and is given by $3 n$.
\subsection{\underline{\textbf{\normalsize{Derivation of energy $E_{\Phi_n} $
in the half-filled case}}}}
The starting state 
$|\phi_n \rangle \equiv |-\vec{k_F} \rightarrow \vec{k}-\vec{p} ,
\vec{k_F} \rightarrow -\vec{k}-\vec{p} ; n_{-\vec{p}}  \rangle $ ($n \ge 1$),
after switching on the interactions, results in the following energy: 
 \begin{eqnarray}
E_{\Phi_n} = T_0+ (n+1) \omega_0
-\sum_{l\neq n}
 \left [ 
\frac{|\langle \phi_l| H_1 |\phi_n\rangle |^2}
{E_{\phi_l}^0 - E_{\phi_n}^0}
 \right ] .
\label{Ephin}
\end{eqnarray}
To evaluate the last term on the right hand side (rhs)
of Eq. (\ref{Ephin}),
we use Eq. (\ref{sec-ord}) and  obtain 
 \begin{eqnarray}
\sum_{l\neq n} 
\frac{|\langle \phi_l| H_1 |\phi_n\rangle |^2}
{E_{\phi_l}^0 - E_{\phi_n}^0}
-{\rm Re} \sum_{\phi_l \neq \psi_n^0}
 \frac{|\langle \phi_l| H_1 |\psi_n^0\rangle |^2}
{E_{\phi_l}^0 - E_{\psi_n^0}^0} 
&&
 = -g^2 \omega_0^2
 {\rm Re}[ \Pi_0 (\vec{k}-\vec{p}, -\vec{k}_F, \omega_0)
\nonumber \\
&&
~~~~~~~~~~~~~~~ +\Pi_0 (-\vec{k}-\vec{p}, \vec{k}_F, \omega_0)]
\nonumber \\
&&
= -2 g^2 \omega_0^2
 {\rm Re} \Pi_0 (\vec{k}-\vec{p}, \vec{k}_F, \omega_0)
  ,
\label{picorr}
\end{eqnarray}
where
 \begin{eqnarray}
{\rm Re} \Pi_0 (\vec{s}, \vec{q}, \omega_0)
\equiv
- \frac{1}{N}
\sum_{\vec{r}}
&&
\left [
\frac{1-n_{\vec{r}}}
{
\epsilon_{\vec{r}} + \omega_0 -
\epsilon_{\vec{s}}}
+
\frac{n_{\vec{r}}}
{
\epsilon_{\vec{q}} + \omega_0 -
\epsilon_{\vec{r}}
}
\right .
\nonumber \\
&&
\left .
~~ -
\frac{n_{\vec{r}}}
{
\epsilon_{\vec{s}} + \omega_0 -
\epsilon_{\vec{r}}
}
-
\frac{1-n_{\vec{r}}}
{
\epsilon_{\vec{r}} + \omega_0 -
\epsilon_{\vec{q}}
}
 \right ] .
\label{pi0}
 \end{eqnarray}
In the above  expression for $\Pi_0$,
the first and second terms correspond to adding
the contributions due to the electron at
$\vec{s}$  going to a state outside the Fermi surface (FS) and those due 
to the electrons within the FS going to the state $ \vec{q}$ respectively.
Whereas the third  and fourth terms, on the rhs of Eq. (\ref{pi0}),
represent subtracting contributions due to electrons within the FS going
to  the state $\vec{s}$ and 
those  due to electron at $ \vec{q}$ going outside the FS respectively.
In obtaining Eq. (\ref{picorr}), the terms that are ignored or overcounted
are negligible for large $N$.
From the main text we know that the state $|\phi_0 \rangle$,
upon turning on the interaction, yields the energy
 \begin{eqnarray}
E_{\Phi_0} = T_0
-\frac{g^2 \omega_0^2}{N}
\sum_{ \vec{q}, m \neq 0 }
\frac{|\langle m|_{el} \rho_{\vec{q}} |0\rangle _{el}|^2}
{ \xi_{m0}+ \omega_0} .
\label{Ephi02}
 \end{eqnarray}
Then, from Eqs. (\ref{Ephin})--(\ref{Ephi02}), we obtain
\begin{eqnarray}
E_{\Phi_n}=
&&
E_{\Phi_0}+
 (n+1)\omega_0 [1+ g^2 \omega_0 {\rm Re}
 \chi_0 (\vec{p}, \omega_0)]
 \nonumber \\
 &&
 + g^2 \omega_0^2
 {\rm Re} [2 \Pi_0 (\vec{k}-\vec{p}, \vec{k}_F, \omega_0)
- \chi_0 (\vec{p}, \omega_0)] .
\label{Ephin22}
\end{eqnarray}
In the above equation, $2\Pi_0 - \chi_0 > 0  $
for all values of $t/\omega_0 > 1$.

\section{}
\begin{figure}
\begin{center}
\includegraphics[width=5.0in]{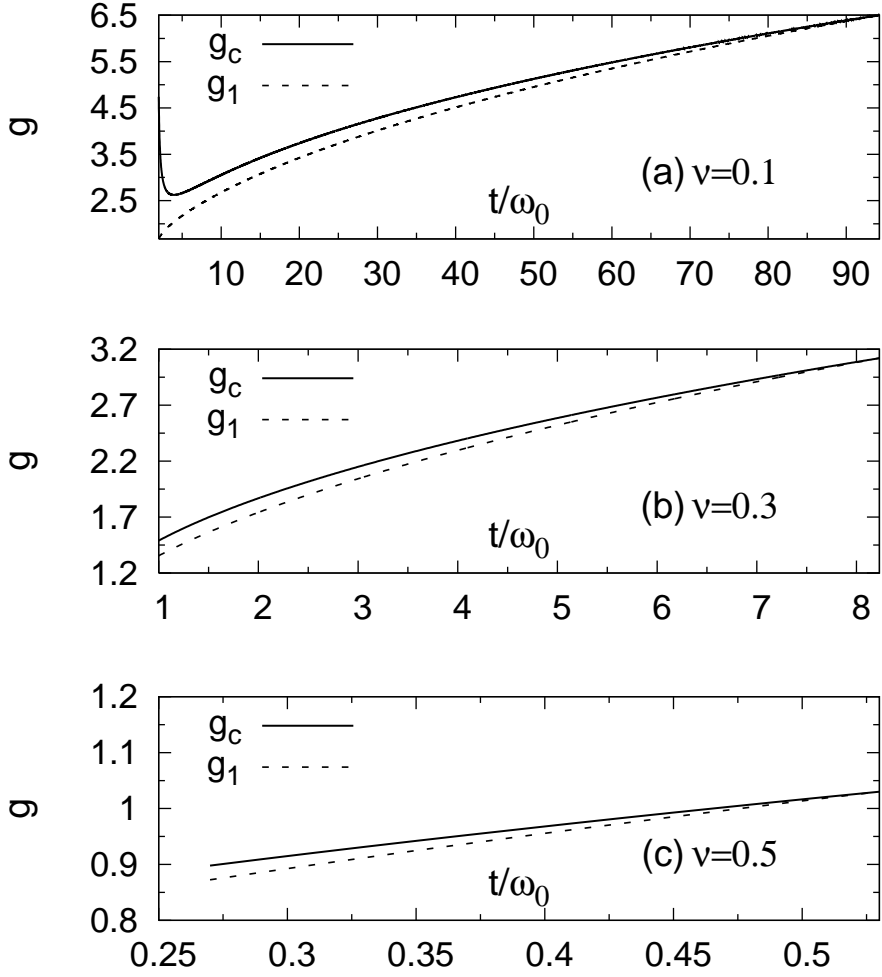}
\end{center}
\vspace*{0.5cm}
\caption[]{Plots of $g_c$ and $g_1$ (drawn till their crossing point)
versus $t/\omega_0$ when the excitation expression
$ \epsilon_{\vec{k}-\vec{p}} - \epsilon_{\vec{k}} = \omega_0$
has only one solution $\vec{k}$ (with $k < k_F=p/2$)
for (a) $\nu = 0.1$ and (b) $\nu=0.3$
 and two solutions for (c) $\nu = 0.5$.}
\label{ggc}
\end{figure}
Here, we consider in detail the non-half-filled case.
For $k_F$ larger (smaller) than $ \pi/4$,
the line given by $t/\omega_0  = 1/[4 \sin(2k_F) \sin(k_F)]$
 corresponds to the larger (smaller)
magnitude wavevector solution $\vec{k}$ lying on the Fermi surface
 as seen from Eq. (\ref{sol2}) (Eq. (\ref{sol1})).
 In the region above (below)
 $t/\omega_0  = 1/[4 \sin(2k_F) \sin(k_F)]$
(see Fig. 3 in the main text), the excitation energy expression
$ \epsilon_{\vec{k}-\vec{p}} - \epsilon_{\vec{k}} = \omega_0$
with $p = 2 k_F$ is satisfied by
one (two) wavevector(s) for $|k| < k_F > \pi/4$ [as seen from
 Eq. (\ref{sol2})] whereas for $|k| < k_F < \pi/4$
it is satisfied by one (zero) wavevector(s) [as seen from Eq. (\ref{sol1})].

In the region $t/\omega_0  >  1/[4 \sin(2k_F) \sin(k_F)]$,
the degenerate states are
$|\psi_n^0 \rangle \equiv |0 ; n_{-\vec{p}} \rangle $ and
$|\psi_n^1 \rangle
 \equiv |\vec{k} \rightarrow \vec{k}-\vec{p} ; (n-1)_{-\vec{p}}
 \rangle $. Unlike at half-filling,
only two (i.e., the above given two) degenerate states need be considered for
carrying out degenerate perturbation theory.
The two basis states are
$|\psi_n^- \rangle \equiv [|\psi_n^0 \rangle - |\psi_n^1 \rangle]/\sqrt{2}$
 and
$|\psi_n^+ \rangle \equiv [|\psi_n^0 \rangle + |\psi_n^1 \rangle]/\sqrt{2}$
with 
$|\psi_n^- \rangle $
($|\psi_n^+ \rangle $)
yielding the lower (higher) eigen energy.
Then, from degenerate perturbation theory, one gets
 \begin{eqnarray}
E_{\Psi_n^-}=
E_{\psi_n^-}
  -\frac{\sqrt{n}}{\sqrt{N}}g\omega_0
-\sum_{\phi_l \neq \psi_n^-, \psi_n^+}
\frac{|\langle \phi_l| H_1 |
\psi_n^- \rangle |^2}
{E_{\phi_l}^0 - E_{
\psi_n^-
}^0} ,
\label{EPsin}
 \end{eqnarray}
 where the second term on the rhs is the first order
energy correction; furthermore, it is understood that
 the states $|\phi_l \rangle$
do not belong to the subspace spanned by $|\psi_n^{\pm} \rangle$. 
Next, to evaluate the last term on the rhs of the above Eq. (\ref{EPsin}),
we use the following non-mixing fact:
 \begin{eqnarray}
\sum_{\phi_l \neq \psi_n^-, \psi_n^+}
\frac{|\langle \phi_l| H_1 |
\psi_n^- \rangle |^2}
{E_{\phi_l}^0 - E_{
\psi_n^-
}^0} =
\frac{1}{2}\sum_{\phi_l \neq \psi_n^-, \psi_n^+}
\frac{|\langle \phi_l| H_1 |\psi_n^0 \rangle |^2}
{E_{\phi_l}^0 - E_{\psi_n^0}^0}
+
\frac{1}{2}\sum_{\phi_l \neq \psi_n^-, \psi_n^+}
\frac{|\langle \phi_l| H_1 | \psi_n ^1\rangle |^2}
{E_{\phi_l}^0 - E_{\psi_n^1}^0} .
\label{mix}
 \end{eqnarray}
Using Eq. (\ref{sec-ord})  and
 on noting that
 \begin{eqnarray}
\sum_{\phi_l \neq \psi_n^-, \psi_n^+}
\frac{|\langle \phi_l| H_1 |\psi_n^1\rangle |^2}
{E_{\phi_l}^0 - E_{\psi_n^1}^0}
=g^2 \omega_0^2
&&
 \left [\frac{1}{N} \right .
\sum_{ \vec{q}, m \neq 0}
\frac{|\langle m|_{el} \rho_{\vec{q}} |0\rangle _{el}|^2}
{ \xi_{m0}+\omega_0}
\nonumber \\
&&
\left .
- (n-1) {\rm Re} \chi_0 (\vec{p}, \omega_0)
-  {\rm Re} \Pi_0 (\vec{k}-\vec{p}, \vec{k}, \omega_0)
\frac{}{}
 \right ] ,
\label{matrix_psin1}
\end{eqnarray}
we obtain from Eqs. (\ref{EPsin})--(\ref{matrix_psin1})
 \begin{eqnarray}
E_{\Psi_n^-}=
&&
E_{\Phi_0}+
E^1
 +n\omega_0 [1+ g^2 \omega_0 {\rm Re}
 \chi_0 (\vec{p}, \omega_0)]
\\
&&
 - 0.5 g^2 \omega_0^2 {\rm Re}
 [\chi_0 (\vec{p}, \omega_0) - \Pi_0 (\vec{k}-\vec{p}, \vec{k}, \omega_0)] .
\label{En1sol2}
\end{eqnarray}
In Eq. (\ref{matrix_psin1}), $\Pi_0 (\vec{k}-\vec{p}, \vec{k}, \omega_0)$ has
 been obtained  in a manner similar to that in the half filled case.
As mentioned in the main text, the macroscopic instability condition
is given by
\begin{eqnarray}
1+ 2 g^2_c \omega_0 {\rm Re}
 \chi_0 (\vec{p}, \omega_0)=0 .
\label{gcchi02}
\end{eqnarray}
However, for each filling below half-filling, there is a
corresponding critical value of the adiabaticity parameter
($t/\omega_0$) below which ${\rm Re} [\chi_0 - \Pi_0] > 0$
 in the above Eq. (\ref{En1sol2}). Consequently,
 the interacting state $|\Psi_1^- \rangle $ produced by
the one phonon state $|\psi_1^- \rangle$
is more stable than the interacting state $|\Phi_0 \rangle $
at an  electron-phonon
coupling value given by $g_{1} < g < g_c$ where $g_1$ corresponds to
$(E_{\Psi_1^-} - E_{\Phi_0})^2 =0$ [see Fig. \ref{ggc}].
Only above this critical adiabaticity parameter value
 do we have $|\Psi_n^- \rangle $ , for $n \rightarrow \infty$,
as the most stable state at a coupling  $g> g_c < g_{1}$.
Furthermore, it should also be noted that the interacting state
$|\Psi_1^- \rangle $ is also a  LL.
Thus, it is obvious that the LL to CDW transition occurs only at
$g = g_c^+$.
Lastly it should also be mentioned that, instead of using the above 
degenerate perturbation theory to obtain the instability condition,
 one can also adopt a blocking approach
similar to that at half-filling by exciting one electron at the Fermi surface
to the state $\vec{k} - \vec{p}$ and then employ
non-degenerate perturbation theory. In the latter case, the condition for
macroscopic instability is still the same while the critical adiabaticity
parameter value (above which $g_c < g_1$) is slightly larger.

\subsection{\underline{\textbf{\normalsize{
The case when
$ \epsilon_{\vec{k}-\vec{p}} - \epsilon_{\vec{k}} 
= \epsilon_{\vec{k}-2\vec{p}} - \epsilon_{\vec{k}-\vec{p}} = \omega_0$}}}}
\begin{figure}
\begin{center}
\includegraphics[width=5.0in]{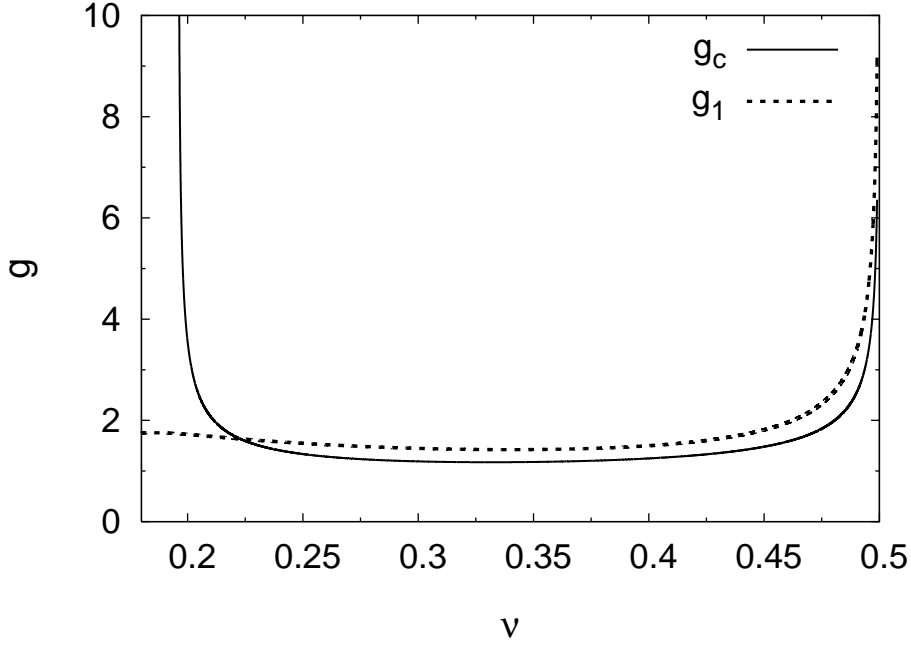}
\end{center}
\vspace*{0.5cm}
\caption[]{Plot of $g_c$ and $g_1$ versus fillings $\nu$  when
 the excitation expressions
$ \epsilon_{\vec{k}-\vec{p}} - \epsilon_{\vec{k}} = \omega_0$
and $ \epsilon_{\vec{k}-2\vec{p}} - \epsilon_{\vec{k}-\vec{p}} = \omega_0$
are simultaneously satisfied by one $\vec{k}$ (with $k < k_F=p/2$).}
\label{k-p+k-2p}
\end{figure}

In the non-half-filled regime, we will now consider the special case where
the excitation expression  
$ \epsilon_{\vec{k}-\vec{p}} - \epsilon_{\vec{k}} = \omega_0$
and $ \epsilon_{\vec{k}-2\vec{p}} - \epsilon_{\vec{k}-\vec{p}} = \omega_0$
are simultaneously satisfied by one $\vec{k}$ (with $k < k_F=p/2$). 
This will hold when $2 t \cos(k) = \omega_0 $
and consequently when $t/\omega_0 = 1/[2\sin(2k_F)]$.
For values of the adiabaticity parameter
$t/\omega_0$ and $k_F$ that lie on the
 line $t/\omega_0 = 1/[2\sin(2k_F)]$ depicted in Fig. 3 of main text,
the method involving only two degenerate states breaks down.
To analyze the CDW instability,
 one can use a blocking method similar to that used at half-filling
in the main text. We Pauli block the
states $\vec{k} -\vec{p}$ and $\vec{k} -2 \vec{p}$ by
the two electrons on the Fermi surface. The blocked state
$|\phi_n^{\prime} \rangle \equiv |\vec{k_F} \rightarrow \vec{k}-\vec{p} ,
-\vec{k_F} \rightarrow \vec{k}-2\vec{p} ; n_{-\vec{p}}  \rangle $
yields the energy for the interacting state $|\Phi_n^{\prime} \rangle $
 to be
 \begin{eqnarray}
E_{\Phi_n^{\prime}}=&&
E_{\Phi_0}+ \omega_0(1+
\csc (k_F))+
 n\omega_0 [1+ g^2 \omega_0 {\rm Re}
 \chi_0 (\vec{p}, \omega_0)] \nonumber \\
&& + g^2 \omega_0^2 {\rm Re}[
  \Pi_0 (\vec{k}-\vec{p}, \vec{k}_F, \omega_0) 
  + \Pi_0 (\vec{k}-2\vec{p}, -\vec{k}_F, \omega_0)
] .
\label{EPhi1}
\end{eqnarray}
Here too the critical coupling $g_c$, for
macroscopic instability, is still given by Eq. (\ref{gcchi02}).
However, for filling factors less than approximately $0.22 $,
 the one phonon interacting state $|\Phi_1^{\prime} \rangle$
is the lowest energy state for $g_1< g < g_c$ (see Fig. \ref{k-p+k-2p}).
At fillings above $0.22 $, the
 large $n$ phonon interacting state $|\Phi_n^{\prime} \rangle$
is the most stable state for $ g > g_c < g_1$.
It should also be mentioned that the above blocking procedure can also
be used when
$ \epsilon_{\vec{k}-\vec{p}} - \epsilon_{\vec{k}} = \omega_0$
and $ \epsilon_{\vec{k}-2\vec{p}} - \epsilon_{\vec{k}-\vec{p}} \approx
\omega_0$ in which case the
 values of $t/\omega_0$ and $k_F$ lie close to the
curve $t/\omega_0 = 1/[2\sin(2k_F)]$.
 
\subsection{\underline{\textbf{\normalsize{
When both the solutions exist}}}}
\begin{figure}
\begin{center}
\includegraphics[width=5.0in]{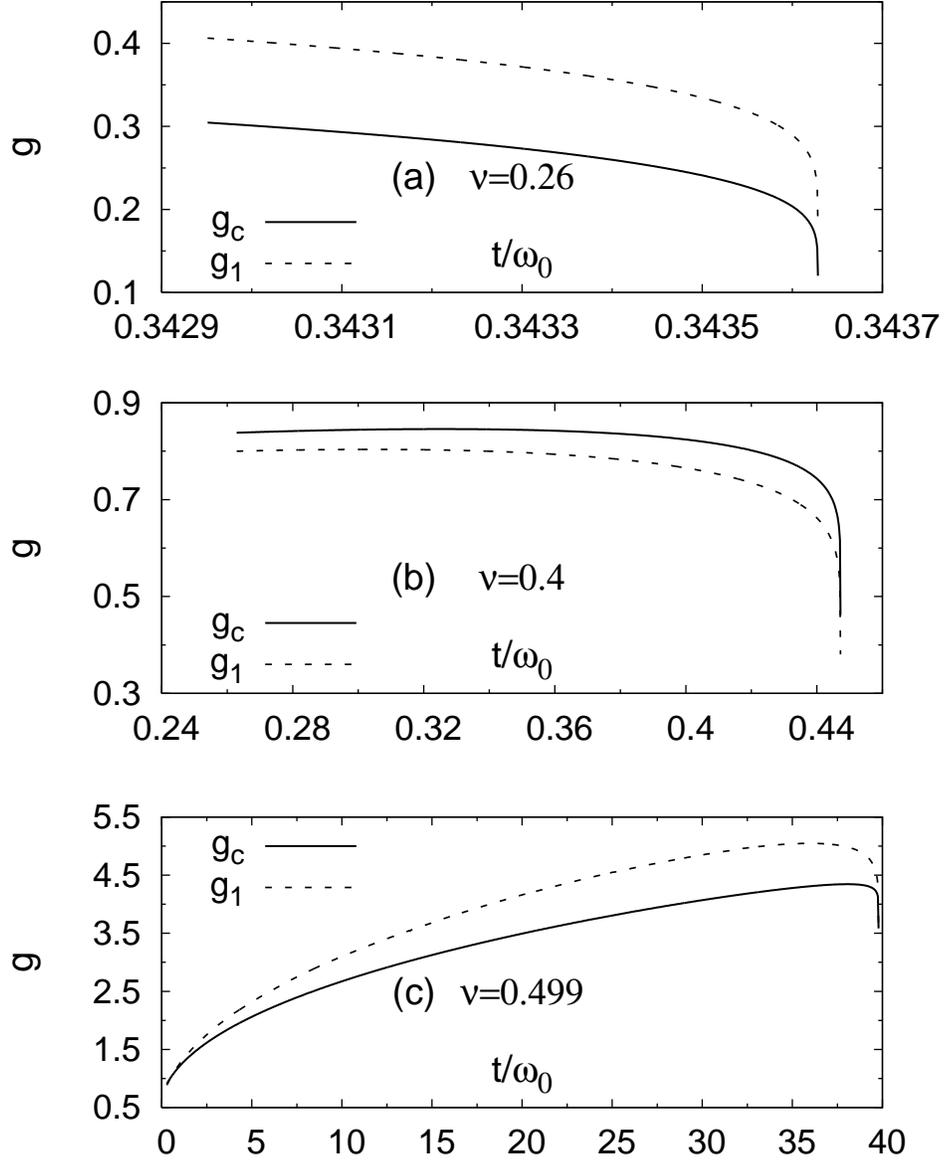}
\end{center}
\vspace*{0.5cm}
\caption[]{Plots of $g_c$ and $g_1$ versus $t/\omega_0$
 when the excitation expression
$ \epsilon_{\vec{k}-\vec{p}} - \epsilon_{\vec{k}} = \omega_0$
has two solutions $\vec{k}$ (with $k < k_F=p/2$)
for (a) $\nu = 0.26$, (b) $\nu=0.4$, and (c) $\nu = 0.499$. }
\label{mlpt2sol}
\end{figure}

Here we will consider the lattice instability for the case 
$t/\omega_0  < 1/[4 \sin(2k_F) \sin(k_F)]$ when two wavevectors satisfy
the excitation energy expression
$ \epsilon_{\vec{k}-\vec{p}} - \epsilon_{\vec{k}} = \omega_0$
 for $k_F > \pi/4$. For $k_F < \pi/4$, only the region
 $t/\omega_0  > 1/[4 \sin(2k_F) \sin(k_F)]$ is relevant as
${\rm Re} \chi_0(\vec{p},\omega_0) =0$ lies above
this line (see Fig. 3 of main text).
For $k_F > \pi/4$, let the two wavevectors that satisfy
$ \epsilon_{\vec{k}-\vec{p}} - \epsilon_{\vec{k}} = \omega_0$
be  $\vec{k}_1$ and $\vec{k}_2$.
We will use the blocking method  to block the
states $\vec{k}_1 -\vec{p}$ and $\vec{k}_2 - \vec{p}$ by
 the two electrons on the Fermi surface. The state
$|\phi_n^{\prime \prime}\rangle \equiv 
|\vec{k_F} \rightarrow \vec{k}_1-\vec{p} ,
-\vec{k_F} \rightarrow \vec{k}_2-\vec{p} ; n_{-\vec{p}}  \rangle $
leads to the energy
 \begin{eqnarray}
E_{\Phi_n^{\prime \prime}}=&&
E_{\Phi_0}+ \omega_0+ 4 t\cos(k_F)+
 n\omega_0 [1+ g^2 \omega_0 {\rm Re}
 \chi_0 (\vec{p}, \omega_0)] \nonumber \\
&& + g^2 \omega_0^2 {\rm Re}[
  \Pi_0 (\vec{k}_1- \vec{p}, \vec{k}_F, \omega_0)
  + \Pi_0 (\vec{k}_2- \vec{p}, -\vec{k}_F, \omega_0)
] .
\label{EPhi2}
\end{eqnarray}
Here also the macroscopic instability occurs when $g > g_c$ with $g_c$ 
obtained from Eq. (\ref{gcchi02}).
For extreme values of $k_F$, i.e., $k_F$ close to $\pi/4$ or $\pi/2$,
 the large $n$ state $|\Phi_n^{\prime \prime} \rangle$
is the lowest energy state for $g > g_c < g_1$ (see Fig. \ref{mlpt2sol}).
On the other hand, at intermediate values of $k_F$,
the state $|\Phi_1^{\prime \prime} \rangle$ is the stable state
for  $ g_1 < g < g_c$ (see Fig.  \ref{mlpt2sol}).
\end{document}